\documentclass[conference]{IEEEtran}
\IEEEoverridecommandlockouts
\usepackage{amsmath,amssymb,amsfonts}
\usepackage{algorithm}
\usepackage{algorithmic}
\usepackage{graphicx}
\usepackage{textcomp}
\usepackage{xcolor}
\def\BibTeX{{\rm B\kern-.05em{\sc i\kern-.025em b}\kern-.08em
    T\kern-.1667em\lower.7ex\hbox{E}\kern-.125emX}}
\usepackage{amsthm}
\usepackage{bm}

\usepackage{multirow}
\usepackage{array}
\usepackage{booktabs}
\usepackage{lipsum}
\usepackage{xcolor}
\usepackage{etoolbox}
\usepackage{algorithm}
\usepackage{algorithmic}
\usepackage{amsthm}

\usepackage{lineno,hyperref}
\modulolinenumbers[5]
\usepackage{bm}
\usepackage{multirow}
\usepackage{array}
\usepackage{booktabs}
\usepackage{lipsum}
\usepackage{xcolor}
\usepackage{etoolbox}
\usepackage{cite}
\ifCLASSINFOpdf
\else
\fi
\ifCLASSOPTIONcompsoc
  \usepackage[caption=false,font=normalsize,labelfont=sf,textfont=sf]{subfig}
\else
  \usepackage[caption=false,font=footnotesize]{subfig}
\fi

\graphicspath{{./FiguresJournal/}} 

\begin{document}

\title{Controlling Grid-Connected Inverters under Time-Varying Voltage Constraints
}
\author{\IEEEauthorblockN{
\IEEEauthorblockA{Zixiao Ma, \IEEEmembership{Member, IEEE}, Baosen Zhang, \IEEEmembership{Member, IEEE}
\\{Department of Electrical and Computer Engineering} \\
{University of Washington}\\
Seattle, WA, USA \\
zixiaoma@uw.edu, zhangbao@uw.edu
}
}
\thanks{This work was supported in part by the Clean Energy Institute at the University of Washington.}
}
\maketitle

\begin{abstract}
Inverter-based resources (IBRs) are becoming increasingly prevalent in power systems. Due to the inherently low inertia of inverters, there is a heightened risk of disruptive voltage oscillations. A particular challenge in the operation of grid connected IBRs is the variations in the grid side voltage. The changes in the grid side voltage introduces nonlinear and time-varying constriants on the inverter voltages themselves. For an operator, it would be useful to know the set of active and reactive powers that can be tracked under these time-varying conditons.  This paper introduces an optimization model designed to assess the achievability of power setpoints within the framework of constrained static state-feedback power control. Additionally, we present a Monte Carlo simulation-based method to optimize the set of achievable power setpoints. The efficacy of the proposed approach is validated through simulation results.
\end{abstract}


\section{Introduction}
In recent years, the growing integration of inverter-based resources (IBRs) into the power grid has become a prominent trend, presenting both opportunities and challenges in grid operations. In this paper, we are interested in the power tracking problem of the IBRs. In particular, we are after controller designs that would quickly converge to desired active and reactive power setpoints, while satisfying the constraints arising from the operation of the IBRs \cite{Chen2009,Zhang2021,Ma2021}.

A key challenge in power tracking arises from the voltage constraints of the IBRs. These constraints require that the inverter output voltage magnitude remains within a safe range centered around the nominal operational point  \cite{Wangzhenji2020}. In contrast to the linear constant voltage constraints found in conventional machines, those on IBR are time-varying with grid voltage and nonlinear concerning the control signals \cite{Ma2023,Du2019,Ma2023b}.

Much work has been done on the power control of IBRs. In direct power control (DPC) for grid-connected voltage source inverters (VSIs), active and reactive powers are regulated without inner-loop current controllers \cite{Hu2011,Noguchi1998,Gui2019b,Gui2023,fu2021hybrid,Gui2018,Gui2018b}. A lookup table-based DPC method was proposed in \cite{Noguchi1998}, which directly selects converter switching signals based on real-time measurements like active power errors and terminal voltage. However, its disadvantage lies in the variable switching frequency, introducing unwanted harmonic spectrum and posing challenges for line filter design. To improve robustness, strategies like sliding mode control and grid voltage modulated DPC were proposed, offering advantages in faster dynamics response \cite{Hu2011,Li2017}. However, the sliding mode control method suffers from the chattering problem and cannot ensure the convergency to the equilibrium. To address these, alternatives like port-controlled Hamiltonian DPC have been explored \cite{Gui2023}. It leverages a dissipative nature, but power ripples persist \cite{Gui2018}.

However, most methods do not address the voltage constraints. They essentially assume that the constraints are always satisfied, which is valid if the system operates in a small region around a feasible point. When the power setpoints and disturbances change more substantially, explicitly addressing the voltage constraints becomes important. One way to do this is model predictive control (MPC) approach \cite{Zhang2017,Peng2023,Hu2015}. It effectively addresses system constraints including control input constraint in DPC of grid-connected VSI. However, challenges arise due to the potential impact of incorrect voltage sequence selection, prompting research into optimal sequences and strategies to mitigate the high computational burden \cite{Gui2018}. Despite these computational approaches, they do not provide guarantees on the achievability. 

To bridge this gap, this paper proposes an optimization model tailored to the particular form of the voltage constraints. It sheds light on attainable power setpoints amidst the interplay of voltage limitations and power generation in grid-connected VSIs. Specially, We express the system linearly with a time-varying nonlinear norm-based constraint, which depends on measurable grid voltage. By leveraging S-lemma, we propose a criterion to assess the achievability of power setpoint achievability. The primary contributions are outlined as follows:
\begin{itemize}
    \item For enhanced computational efficiency, we adopt a static state-feedback controller to circumvent the time-consuming iterative online optimization required in MPC.
    \item We propose a systematic method for evaluating the achievability of power setpoints. This ensures the generation of a secure control signal that adheres to constraints on VSI output voltage magnitude while accounting for time-varying grid voltage.
    \item By introducing a Monte Carlo simulation-based method, we optimize the set of achievable power setpoints. This approach offers a practical solution for navigating complex decision landscapes faced by system operators.

\end{itemize}

\section{Problem Statement}
In this section, we first introduce the model of grid-connected VSI. Then, Then we show how the time-varying constraints on the input arise. Finally, the overall power control problem of grid-connected VSI is presented.
\subsection{VSI Modeling}
Considering a three-phase balanced voltage condition, the dynamics of instantaneous active power $P$ and reactive power $Q$  can be represented in the $\alpha-\beta$ stationary reference frame as \cite{Gui2018}:
\begin{subequations}\label{system_pq}
   \begin{align}
    \dot{P}&=-\frac{R}{L}P-\omega Q+\frac{3}{2L}\left(v_{{\rm G}\alpha}u_{\alpha}+v_{{\rm G}\beta}u_{\beta}-V_{\rm G}^2\right),\\
    \dot{Q}&=\omega P-\frac{R}{L} Q+\frac{3}{2L}\left(v_{{\rm G}\beta}u_{\alpha}-v_{{\rm G}\alpha}u_{\beta}\right),
\end{align} 
\end{subequations}
where $R$ and $L$ are filter resistance and inductance, respectively; $\omega=2\pi f$ is the angular frequency and $f$ is the frequency of the grid; $u_{\alpha}$ and $u_{\beta}$ are VSI output voltages in $\alpha-\beta$ coordinates. The voltage magnitude of the grid denoted as $V_{\rm G}$ is time-varying. It enters (\ref{system_pq}) in the $\alpha-\beta$ coordinates as $v_{{\rm G}\alpha}$ and $v_{{\rm G}\beta}$, where
\begin{subequations}\label{VG}
\begin{align}
    v_{{\rm G}\alpha}&=V_{\rm G}\cos{\omega t},\\
    v_{{\rm G}\beta}&=V_{\rm G}\sin{\omega t}.
\end{align}
\end{subequations}

The state of system (\ref{system_pq}) is $\mathbf{x}=[P\;\;Q]^{\top}$. It is possible to take $u_\alpha$ and $u_\beta$ as the control inputs, but it will lead to a viarant coefficient depending on $v_{{\rm G}\alpha}$ and $v_{{\rm G}\beta}$. To simplify the controller design, we define the control input as follows
\begin{align*}
   \mathbf{u}=\begin{bmatrix}
        u_P\\
        u_Q
    \end{bmatrix}=\begin{bmatrix}
        v_{{\rm G}\alpha}u_{\alpha}+v_{{\rm G}\beta}u_{\beta}\\
        v_{{\rm G}\beta}u_{\alpha}-v_{{\rm G}\alpha}u_{\beta}
    \end{bmatrix}.
\end{align*}
Since $V_{\rm G}^2$ is time-varying, we think of $V_{\rm G}^2$ as the additive disturbance into the system, and the system matrices are
\begin{align*}
    A=\begin{bmatrix}
       -\frac{R}{L} & -\omega\\\omega&-\frac{R}{L}
    \end{bmatrix}, \;B=\begin{bmatrix}
        \frac{3}{2L}&0\\0& \frac{3}{2L}
    \end{bmatrix},\;\;E=\begin{bmatrix}
        -\frac{3}{2L}\\ 0
    \end{bmatrix}.
\end{align*}
Then, the dynamics can be summarized as
\begin{align}\label{system}
    \dot{\mathbf{x}}=A\mathbf{x}+B\mathbf{u}+EV_{\rm G}^2.
\end{align}
Note that the physical control signals of the inverter can be recovered can be recovered from $\mathbf{u}$ as
\begin{align}
    \begin{bmatrix}
        u_{\alpha}\\u_{\beta}
    \end{bmatrix}&=\begin{bmatrix}
          v_{{\rm G}\alpha}&v_{{\rm G}\beta}\\
        v_{{\rm G}\beta}&-v_{{\rm G}\alpha}
    \end{bmatrix}^{-1}\begin{bmatrix}
        u_P\\
        u_Q
    \end{bmatrix}\\
    &=\frac{\|\mathbf{u}\|_2}{V_{\rm G}}\begin{bmatrix}
        \cos{\omega t}&\sin{\omega t}\\
        \sin{\omega t}&-\cos{\omega t}
    \end{bmatrix}.
\end{align}
\subsection{Input Constraint Development}
\begin{figure}[ht!]
		\centering
		\includegraphics[width=0.95\columnwidth]{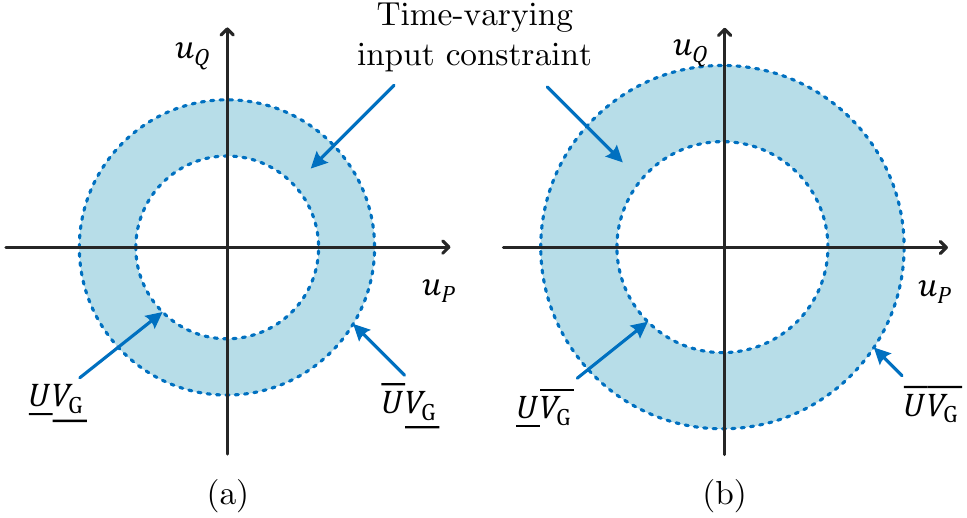}
		\caption{The input constraint denoted as the blue area is nonconvex and the exact achievable controls depend on values of $V_{\rm G}$, as shown in (a) and (b). }
		\label{inputconstraintfig}
						\vspace{-1 em}
	\end{figure}

In practice, the VSI output voltage magnitude $U$ should be bounded around the rated output voltage, i.e.,
\begin{align}
    \underline{U}\leqslant U\leqslant\overline{U}
\end{align}
where $\underline{U}$ and $\overline{U}$ are the lower and upper bounds of the VSI output voltage magnitude.
Since $\|\mathbf{u}\|_2=\sqrt{u_P^2+u_Q^2}=UV_{\rm G}$, the input constraint can be written as
\begin{align}\label{inputconstraint}
   \underline{U}V_{\rm G}\leqslant \|\mathbf{u}\|_2\leqslant \overline{U}V_{\rm G}.
\end{align}
Even though the grid voltage is time-varying, it is bounded. This creates a bound on the input and the disturbance, i.e.,
\begin{align}\label{dconstraint}
    \underline{V_{\rm G}}\leqslant V_{\rm G}\leqslant\overline{V_{\rm G}},
\end{align}
where $\underline{V_{\rm G}}$ and $\overline{V_{\rm G}}$ are lower and upper bounds of the grid voltage, respectively. The form of (\ref{inputconstraint}) implies that the constraint on the input is nonconvex and time-varying, as shown in Fig. \ref{inputconstraintfig}. These types of constraints on the input is somewhat unique to inverter control, and many existing control design techniques do not readily apply.
\subsection{The Power Control Problem}
\begin{figure}[t!]
		\centering
		\includegraphics[width=0.95\columnwidth]{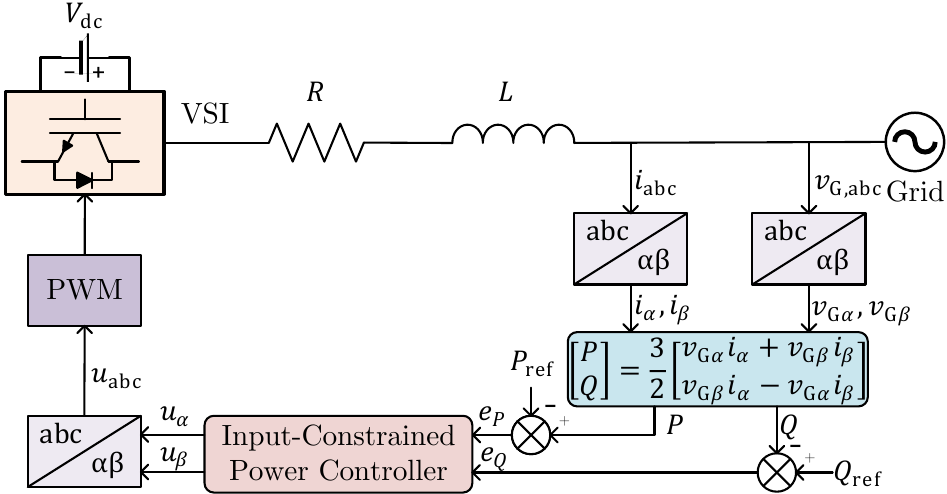}
		\caption{Control block diagram of grid-connected VSI for power regulation.}
		\label{systemfig}
						\vspace{-1.2 em}
	\end{figure}

As depicted in Fig. \ref{systemfig}, the primary control objective is to effectively regulate the VSI output active and reactive powers to align with their designated setpoints, denoted as $\mathbf{x}_{\rm ref}=\left[P_{\rm ref}\;\;Q_{\rm ref}\right]^\top$. This regulation is subject to the time-varying input constraint (\ref{inputconstraint}). Because of this constraint, not all possible setpoints $\mathbf{x}_{\rm ref}$ can be achieved. In fact, it is trivial to see if a particular given setpoint is achievable. We say a setpoint $\mathbf{x}_{\rm ref}$ is achievable if there exists control actions $\mathbf{u}$, where $\mathbf{u}$ satisfies input constraint (\ref{inputconstraint}) and $\mathbf{x}(t)\to \mathbf{x}_{\rm ref}$.  Given a controller $\mathbf{u}$, we call the set of all achievable setpoints the achievable set. Consequently, two natural questions arise: 1) Given a control algorithm, what is the achievable set? 2) Can we design the controller to maximize the area of this set? We attempt to answer these questions in the rest of the paper. 
\section{Input-Constrained Power Controller Design and Achievable Set Optimization}
This section presents how to design an input-constrained power controller. We use S-lemma to assess the achievability of power setpoints, which is formulated as a semi-definite programming (SDP) problem. Subsequently, we design a Monte Carlo simulation to optimize the size of the achievable set.
\subsection{Power Controller Design}
System (\ref{system}) is controllable and the time-varying disturbance $V_{\rm G}^2$ is measurable. Therefore, the active and reactive power can be regulated to the desired setpoints with a simple feedforward controller $\mathbf{u}=-B^{-1}A\mathbf{x}_{\rm ref}-B^{-1}EV_{\rm G}^2$. 

To optimize the area of the achievable set as well as enable the adjustment of control performance, we design a static state-feedback power controller as follows:
\begin{align}\label{controller}
    \mathbf{u}=-K(\mathbf{x}-\mathbf{x}_{\rm ref})-B^{-1}A\mathbf{x}_{\rm ref}-B^{-1}EV_{\rm G}^2,
\end{align}
where $K \in \mathbb{R}^{2\times 2}$ is feedback gain to be optimized to maximize the size of achievable set. Defining error state as $\mathbf{e}=\mathbf{x}-\mathbf{x}_{\rm ref}$, the closed-loop error dynamics can be obtained by substituting (\ref{controller}) into (\ref{system}):
\begin{align}\label{errordynamics}
    \dot{\mathbf{e}}=\left(A-BK\right)\mathbf{e}.
\end{align}
Thus, $\mathbf{x}$ tends to $\mathbf{x}_{\rm ref}$ exponentially fast, if and only if all the eigenvalues of $A-BK$ have strictly negative real parts.
\subsection{Criterion of Setpoint Achievability}
We aim to find a set of setpoints such that (\ref{inputconstraint}) holds for all time under (\ref{dconstraint}). To simplify the derivation, we define $d=V_{\rm G}^2$, $\underline{d}=\underline{V_{\rm G}}^2$, and $\overline{d}=\overline{V_{\rm G}}^2$, restating the problem as:
\begin{align}\label{problem_o}
    \underline{d}\leqslant d\leqslant\overline{d}
    \;\Rightarrow\;\underline{U}\sqrt{d}\leqslant \|\mathbf{u}\|_2\leqslant \overline{U}\sqrt{d}\;\;\;\forall t.
\end{align}

The above problem is nontrivial because the input constraint is nonconvex. Since (\ref{inputconstraint}) is composed of two quadratic inequalities, a commonly used tool is the S-lemma. This lemma determines when one quadratic inequality implies another. To use it, we first convert (\ref{dconstraint}) to a quadratic inequality:
\begin{align}
    \mathcal{Q}_a=-d^2+(\underline{d}+\overline{d})d-\underline{d}\overline{d}\geqslant 0.
\end{align}
It is straightforward to check that $\mathcal{Q}_a\geqslant 0$ is equivalent to inequality (\ref{dconstraint}). To cope with state variables, we define 
\begin{align}
    \mathbf{z}=\left[d\;P\;Q\right]^{\top},\; M=\left[B^{-1}E\;\;K\right]\in\mathbb{R}^{2\times3},
\end{align}
such that $\mathcal{Q}_a$ can be equivalently rewritten as
\begin{align}\label{qa}
    \mathcal{Q}_{a}:\;\;\mathbf{z}^{\top}{Q_a}\mathbf{z}+\mathbf{r}_{a}^{\top}\mathbf{z}+{c_a}
\geqslant0,
\end{align}
where
\begin{align*}
   Q_a=\begin{bmatrix}
    -1 &&\\&0&\\&&0
    \end{bmatrix},\;\mathbf{r}_{a}=\begin{bmatrix}
    \underline{d}+\overline{d}&0&0
    \end{bmatrix}^{\top},\; c_a=-\underline{d}\overline{d}
\geqslant0.
\end{align*}

Then, we substitute (\ref{controller}) into  (\ref{inputconstraint}) and obtain 
\begin{align}\label{inputconstraint2}
   \underline{U}\sqrt{d}\leqslant \|(K-B^{-1}A)\mathbf{x}_{\rm ref}-M\mathbf{z}\|_2\leqslant \overline{U}\sqrt{d}.
\end{align}
By defining $\underline{\mathbf{h}}=[\underline{U}^2\;0\;0]^{\top}$ and $\overline{\mathbf{h}}=[\overline{U}^2\;0\;0]^{\top}$, we break (\ref{inputconstraint2}) into two quadratics:
\begin{subequations}\label{qb}
    \begin{align}
    \mathcal{Q}_{b1}=\mathbf{z}^{\top}{Q_{b1}}\mathbf{z}+\mathbf{r}_{b1}^{\top}\mathbf{z}+{c_{b1}}\geqslant0 \label{qb1},\\
    \mathcal{Q}_{b2}=\mathbf{z}^{\top}{Q_{b2}}\mathbf{z}+\mathbf{r}_{b2}^{\top}\mathbf{z}+{c_{b2}}\geqslant0, \label{qb2}
\end{align}
\end{subequations}
where 
\begin{align*}
   {Q_{b1}}&=M^{\top}M,\;{\mathbf{r}_{b1}}=-2[(K-B^{-1}A)\mathbf{x}_{\rm ref}]^{\top}M-\underline{\mathbf{h}}^{\top},\\ {c_{b1}}&=[(K-B^{-1}A)\mathbf{x}_{\rm ref}]^{\top}[(K-B^{-1}A)\mathbf{x}_{\rm ref}],\\
   {Q_{b2}}&=-M^{\top}M,\;{\mathbf{r}_{b2}}=2[(K-B^{-1}A)\mathbf{x}_{\rm ref}]^{\top}M+\overline{\mathbf{h}}^{\top},\\ {c_{b2}}&=-[(K-B^{-1}A)\mathbf{x}_{\rm ref}]^{\top}[(K-B^{-1}A)\mathbf{x}_{\rm ref}].
\end{align*}

Finally, problem (\ref{problem_o}) becomes finding values of $\mathbf{x}_{\rm ref}$ such that (\ref{qa})$\Rightarrow$ (\ref{qb}). According to the S-lemma \cite{Polik2007}, $\mathcal{Q}_{a}\geqslant0\Rightarrow \mathcal{Q}_{bi}\geqslant0$, if and only if there exists a $\lambda_i$ such that 
\begin{align*}
    \mathcal{Q}_{bi}-\lambda_i\mathcal{Q}_{a}\geqslant0,\;\;i=1,2.
\end{align*}
This can be converted to the following SDP problems:
\begin{subequations}
\begin{align}
    {\rm P1}:\;&\min_{\lambda_1} \lambda_1^2\\
    {\rm s.t.}\; &\begin{bmatrix}
        Q_{b1}-\lambda_1Q_a&\frac{1}{2}\left(\mathbf{r}_{b1}-\lambda_1\mathbf{r}_{a}\right)\\\frac{1}{2}\left(\mathbf{r}_{b1}^{\top}-\lambda_1\mathbf{r}_{a}^{\top}\right)&c_{b1}-\lambda_1c_a
    \end{bmatrix}\geqslant0,
\end{align}
\end{subequations}
\begin{subequations}
\begin{align}
    {\rm P2}:\;&\min_{\lambda_2} \lambda_2^2\\
    {\rm s.t.}\; &\begin{bmatrix}
        Q_{b2}-\lambda_2Q_a&\frac{1}{2}\left(\mathbf{r}_{b2}-\lambda_2\mathbf{r}_{a}\right)\\\frac{1}{2}\left(\mathbf{r}_{b2}^{\top}-\lambda_2\mathbf{r}_{a}^{\top}\right)&c_{b2}-\lambda_2c_a
    \end{bmatrix}\geqslant0.
\end{align}
\end{subequations}

Given a static feedback gain $K$ such that the matrix $A-BK$ is stable, the set of achievable values of $\mathbf{x}_{\rm ref}$ is characterized by the simultaneous achievability of both SDP problems ${\rm P1}$ and ${\rm P2}$. Within this achievable set, the satisfaction of the input constraint (\ref{inputconstraint}) is guaranteed throughout the entire duration.
\subsection{Optimize the Achievable Set}
\begin{figure}[t!]
		\centering
		\includegraphics[width=0.79\columnwidth]{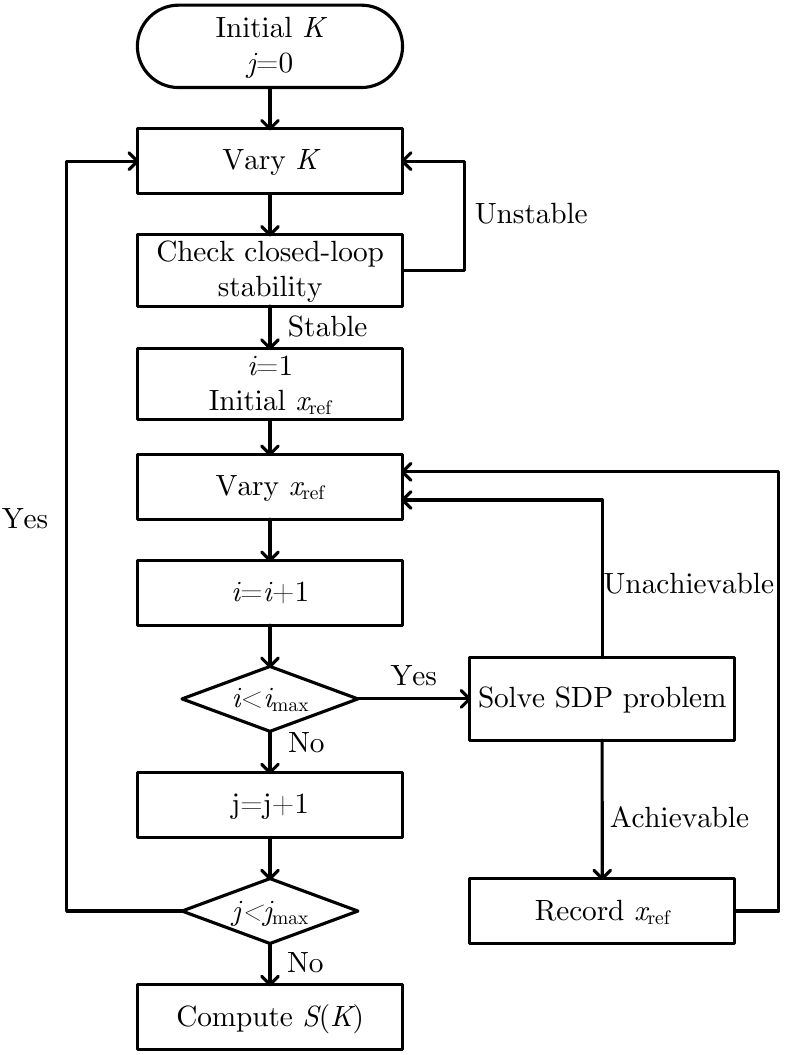}
		\caption{Flowchart of the optimization algorithm for maximizing the achievability set based on Monte Carlo simulation. $i_{\rm max}$ and $j_{\rm max}$ are the total numbers of sampled $\mathbf{x}_{\rm ref}$ and $K$, respectively.}
		\label{algorithm}
	\end{figure}
Introducing feedback gain $K$ in our proposed controller allows for optimizing the area of the achievable set. It turns out that the shape of the achievable sets does not admit closed-form formula for their area. Therefore, we take a statistical approach to approximate its area. We sample uniformly at random in a domain, and we define achievability rate as a metric:
\begin{align}
    S(K)=\frac{\text{number of achievable}\; \mathbf{x_{\rm ref}}}{\text{number of all sampled} \;\mathbf{x_{\rm ref}}}.
\end{align}

Based on the achievability outcomes of the SDP problems, we compute $S(K)$. This mapping facilitates the maximization of $S(K)$ by tuning $K$ through simulation-based optimization. Notably, given that $K$ comprises only four elements, a straightforward Monte Carlo simulation is effective in this scenario. The developed optimization algorithm is depicted in Fig. \ref{algorithm}.

\section{Case studies}
In this section, we conduct case studies to validate the performances of the proposed input-constrained power controller and achievability optimization model using the MATLAB environment. The test system is illustrated in Fig. \ref{systemfig}, and the system parameters are specified in Table \ref{parameters}. In the Monte Carlo simulation, the four elements $(k_{11},k_{12},k_{21},k_{22})$ in $K$ are uniformly sampled within a neighborhood around zero. To optimize $S(K)$, we iterate through all the samples.

\begin{figure}[t!]
		\centering
		\includegraphics[width=0.9\columnwidth]{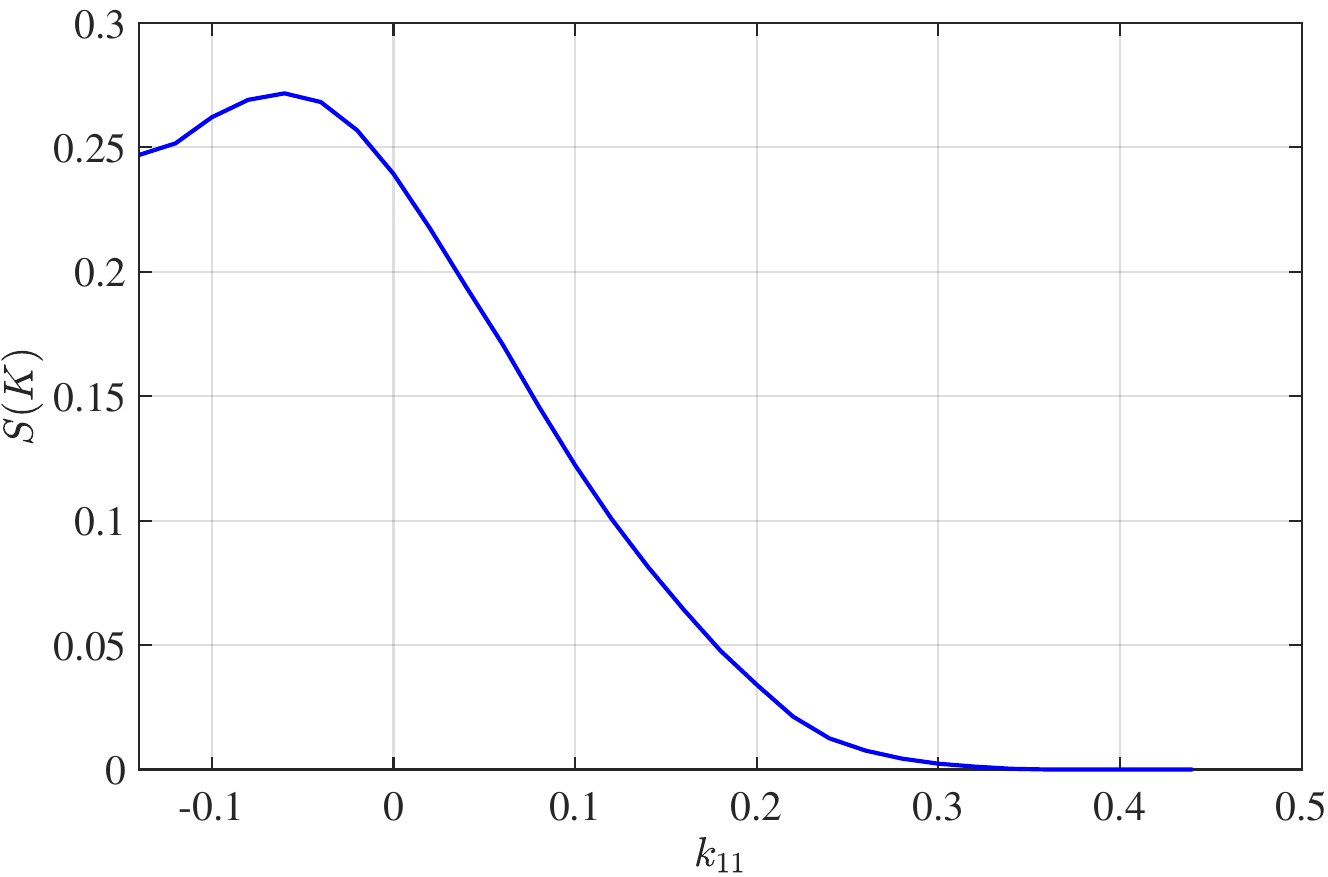}
		\caption{Achievability rates $S(K)$ with respect to the $k_{11}$ coordinate of $K$ in the Monte Carlo simulation.}
		\label{feasirate}
	\end{figure}

\begin{table}[t!]
\setlength{\tabcolsep}{5pt}
\centering
\caption{Parameter setting of grid-connected VSI}
\footnotesize
\renewcommand\arraystretch{1.5}
\label{parameters}
\begin{tabular}{p{1cm}<{\raggedright}p{2.5cm}<{\raggedright}p{1cm}<{\raggedright}p{1cm}<{\raggedright}}
\hline\hline
 Par. & Value  & Par. & Value \\ \hline
 $R$ & $0.12 \;\Omega$ & $L$ & $4$ mH\\
 $\underline{V_{\rm G}}$ & $105.6$ V & $\overline{V_{\rm G}}$ & $114.4$ V\\
  $\underline{U}$ & $104.5$ V & $\overline{U}$ & $115.5$ V\\
  $\omega$ & $314$ rad/s & $f$ & $50$ Hz
	\\ \hline \hline
		\end{tabular}
				\vspace{-1 em}
	\end{table}
As an example to show the Monte Carlo simulation result, the achievability rate $S(K)$ concerning the first coordinate $k_{11}$ of $K$ is shown in Fig. \ref{feasirate}. The optimized $K=\begin{bmatrix}
    -0.08&-0.06\\0.02&-0.16
\end{bmatrix}$, resulting in a $33\%$ improvement of achievability rate $S(K)$ compared to the open-loop case, i.e., $K=\mathbf{0}$. 

\begin{figure}[t!]
		\centering
		\includegraphics[width=0.95\columnwidth]{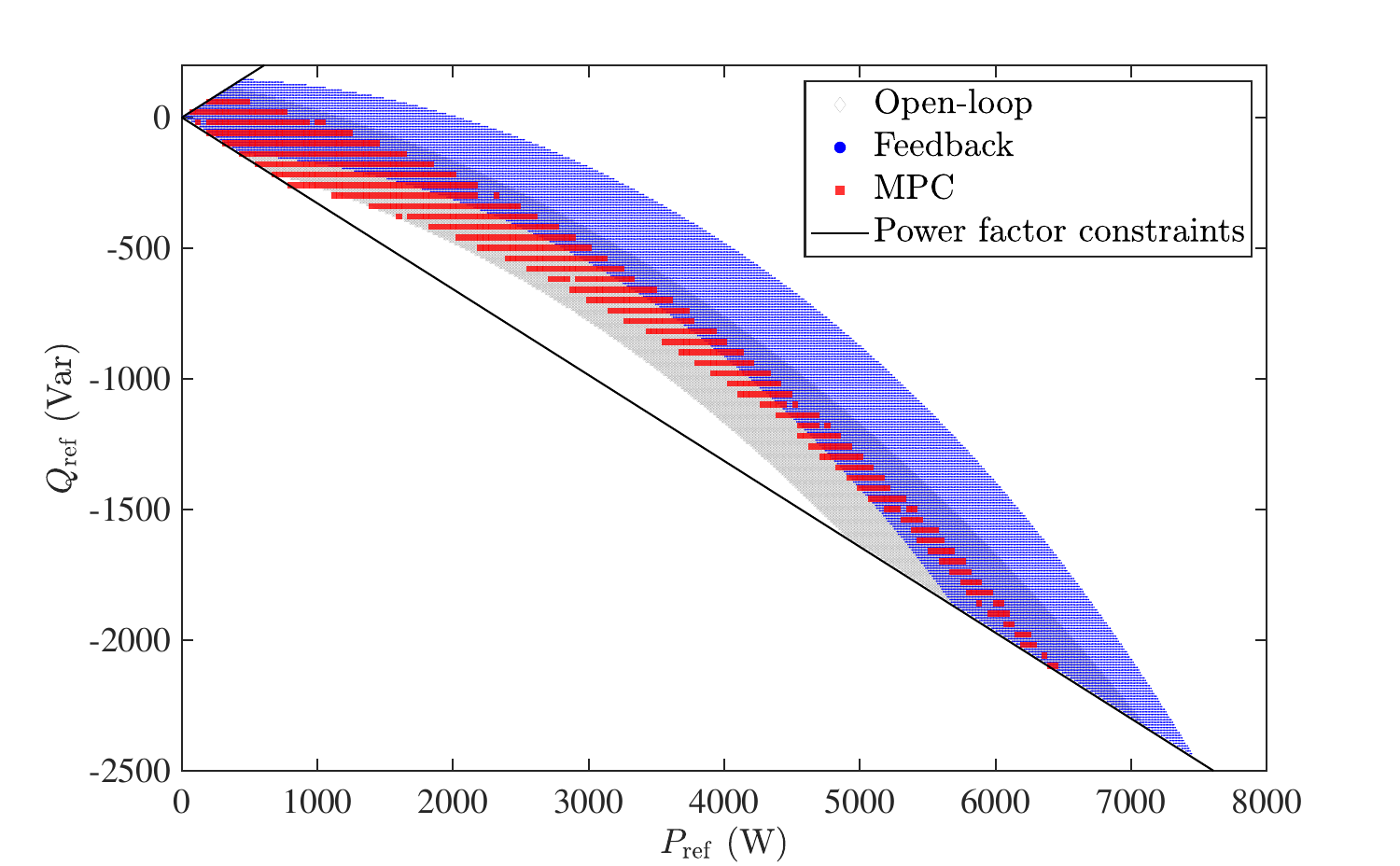}
		\caption{The achievable regions of setpoints with respect to the open-loop controller, feedback controller and MPC. The feedback controller enables optimizing the achievable region by tuning $K$, thus yielding the largest area.}
		\label{feasiregion}
	\end{figure}
To evaluate the achievable region, $P_{\rm ref}$ is sampled uniformly within $[0,8000]$ W and $Q_{\rm ref}$ is sampled uniformly within $[-2500,200]$ Var, while ensuring the power factor with $[0.95,1]$. We compare the achievable sets obtained by open-loop, feedback, and MPC methods. Since the future $V_{\rm G}$ is unavailable, we consider $\underline{U}\overline{V_{\rm G}}$ and $\overline{U}\underline{V_{\rm G}}$ as the lower and upper constraint bounds in the prediction stage of MPC, respectively, to ensure satisfaction of (\ref{inputconstraint}) for all time. As shown in Fig. \ref{feasiregion}, the proposed optimization method can effectively enlarge the achievable region compared with the open-loop controller and MPC. In addition, the computational time for achievable sets of our method and MPC are 26 s and 21712 s, respectively, showing the better efficiency of the proposed method.
 \begin{figure}[t!]
		\centering
		\includegraphics[width=0.95\columnwidth]{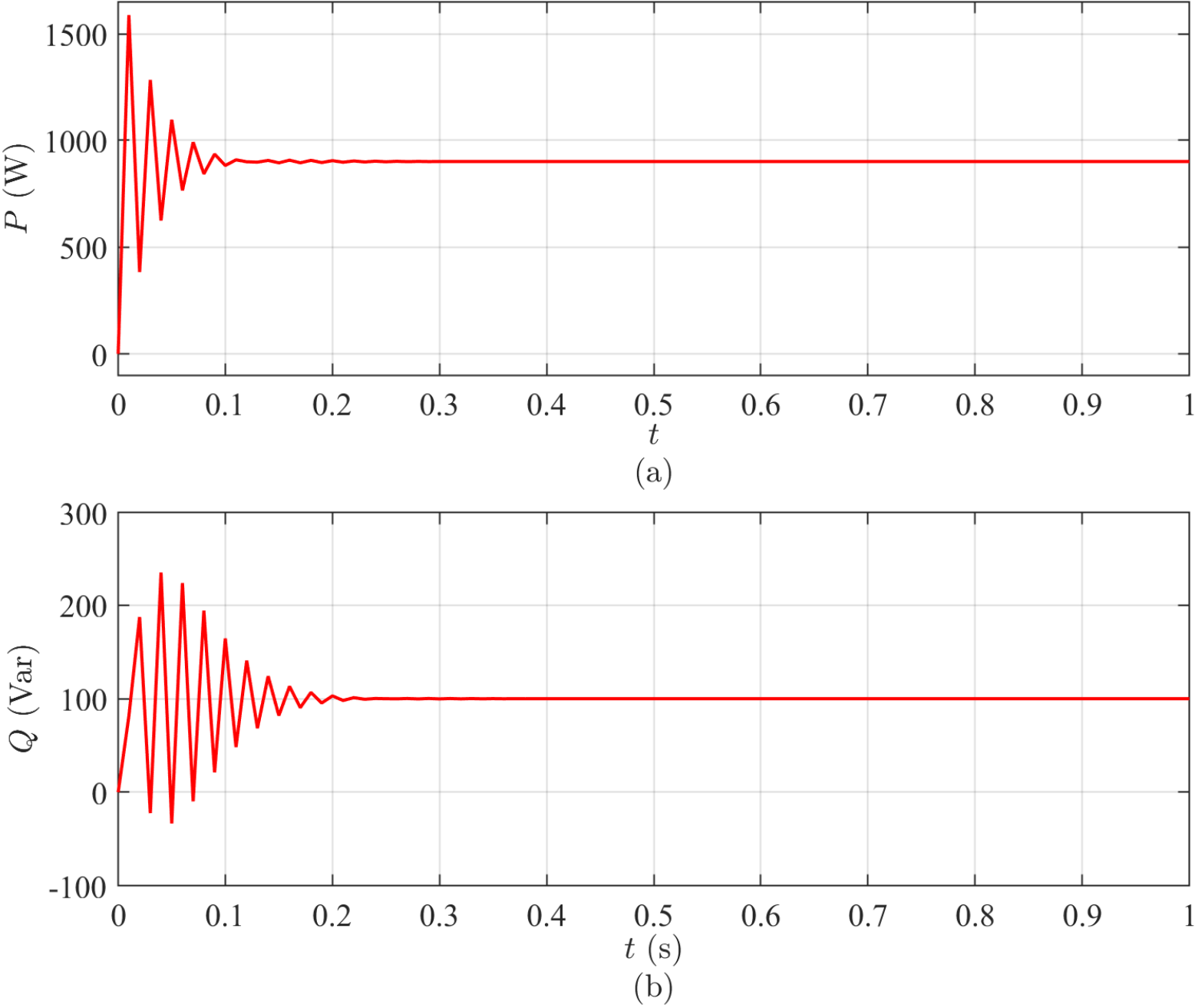}
		\caption{Control performance of the optimized $K$. The initial point is the origin and the setpoint is $\mathbf{x}_{\rm ref}=[900\;\;100]^{\top}$.}
		\label{response}
	\end{figure}

\begin{figure}[t!]
		\centering
		\includegraphics[width=0.95\columnwidth]{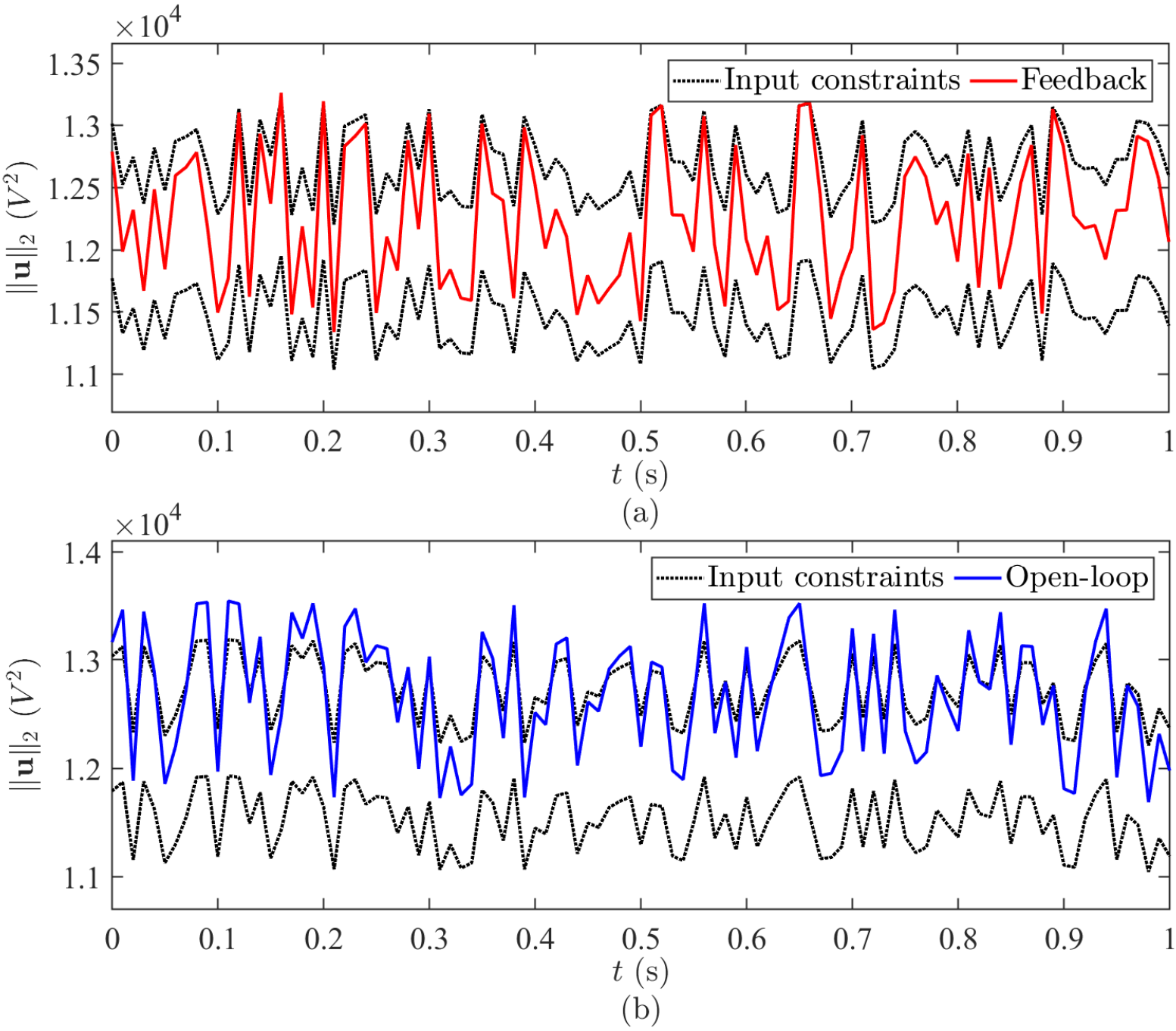}
		\caption{Control signals with optimized $K$: (a) the setpoint is inside the achievable region; (b) the setpoint is outside the achievable region. The red curve stays within the time-varying input constraints marked by the black dotted curves, while the blue curve occasionally exceeds the boundary.}
		\label{inputconstraint_sim}
	\end{figure}

The control performance of the optimized $K$ is shown in Fig. \ref{response}. To assess compliance with input constraints, we maintain $K$ at its optimized value and choose two setpoints—one within the achievable region ($\mathbf{x}_{\rm ref}=[900\;\;100]^{\top}$) and another outside it ($\mathbf{x}_{\rm ref}=[1200\;\;300]^{\top}$) as depicted in the blue region in Fig. \ref{feasiregion}. As illustrated in Fig. \ref{inputconstraint_sim}, the control signal corresponding to achievable setpoint consistently remains within the time-varying input constraint, whereas the signal associated with unachievable setpoint may exceed the boundaries.
\section{Conclusion}
This paper proposes an optimization model for evaluating the achievability of power setpoints within the constraints of static state-feedback power control. Through the integration of a static state-feedback controller for computational efficiency and a systematic method to assess setpoint achievability, our approach aims to elucidate the intricate relationship between voltage source inverter (VSI) output voltage boundedness and its power generation capacity. Furthermore, the proposed Monte Carlo simulation-based method offers a pragmatic solution for system operators, enabling a judicious balance between control performance and the size of the achievable set under the constraints of control inputs. Simulation results affirm the efficacy of our strategy in simultaneously satisfying constraints and optimizing achievable regions. 
\bibliographystyle{IEEEtran}
	\bibliography{reference}
\end{document}